\RequirePackage{fix-cm}
\documentclass[smallextended]{svjour3}       
\smartqed  
\usepackage{graphicx}
\begin{document}

\title{UV lines as a tracers for the XUV-fluxes of stars and the PLATOspec project}
\titlerunning{PLATOspec}        
\author{Eike Wolf Guenther \and Petr Kabath \and Leonardo Vanzi}
\authorrunning{E.W. Guenther} 
\institute{E.W. Guenther \at
              Th\"uringer Landessterwarte Tautenburg, 
              07778 Tautenburg,  Germany \\
              Tel.: +49-36427-86355\\
              Fax: +49-36427-86329\\
              \email{guenther@tls-tautenburg.de}           
           \and
            Petr Kab\'{a}th \at
            Astronomical Institute of the Czech Academy of Sciences, 
            Fri\v{c}ova 298, 25166, Ond\v{r}ejov, Czech Republic
           \and 
            Leonardo Vanzi \at
            Centre of Astro-Engineering, Pontificia Universidad Catolica de Chile, 
            Av. Vicu\~na Mackenna 4860, Santiago, Chile, and
            Department of Electrical Engineering, Pontificia Universidad 
            Catolica de Chile, Av. Vicu\~na Mackenna 4860, Santiago, Chile}

\date{Received: date / Accepted: date}

\maketitle

\begin{abstract}

Observations in the UV-regime are very important for exoplanet
research, because many diagnostically important lines for studying
stellar activity are in this regime. Studying stellar activity is not
only important because of its negative effects on the determination
planetary parameters, but also because the XUV-radiation from the host
stars affects the photochemistry and the erosion of planetary
atmospheres . Unfortunately, the XUV-region is only accessible from
space. However, since the XUV-radiation is correlated with the
Ca\,{\tiny II}\,H\&K-lines, we can use these lines to study the XUV
radiation indirectly. The Ca\,{\tiny II}\,H\&K-lines for relatively
bright stars can be observed with PLATOspec, a new high-resolution
echelle spectrograph in development for the ESO 1.5m telescope at La
Silla. One advantage compared to instruments on larger telescopes will
be that large programs can be carried out.  There will be two modes
for obtaining precise RV-measurements.  In the future, the CUBES
instrument on the VLT will be able to study the same lines to probe
the XUV-radiation in much fainter targets.  \keywords{extrasolar
  planet \and stellar activity \and CaII,HK-lines \and PLATO \and
  ARIEL}


\end{abstract}

\section{The next generation of exoplanet missions: PLATO and ARIEL}
\label{sec:1}

The discovery of the first exoplanet around a Sun-like star, 51 Peg b
opened up a totally new field of research in astronomy
\cite{Mayor1995}.  The radial-velocity (RV) method used for this
discovery initially allows to determine only $\rm M_{p,min}=M_p\times
sin\,i$ ($M_p$ mass of the planet and i, the inclination of its
orbit).  If we observe many stars, we obtain a statistical mass
distribution of the whole ensemble.  Statistically, the true mass of a
planet is $\rm M_p \sim 4/\pi\times M_{p,min}$.  Complementary to the
RV-method is the transit-method, which allows to determine the radii
and inclinations of the planets.  Combining the two allows to
determine the true masses, radii and hence the densities of
exoplanets.  Density measurements are currently the best way to
constrain the composition of exoplanets. Gas-giants in our
solar-system have densities between 0.7 and 1.6 $\rm g\,cm^{-3}$,
rocky planets between 3.9 and 5.5 $\rm g\,cm^{-3}$.  Exoplanets have
been found that have densities as low as 0.7 $\rm g\,cm^{-3}$
(e.g. Kepler 51) and higher than 5.5 $\rm g\,cm^{-3}$ (e.g. K2-106 b,
K2-229 b and 107c).  There are also exoplanets with intermediate
densities in the range between 1.6 and 3.9 $\rm g\,cm^{-3}$.
Exoplanets are turn out to be much more divers than the planets in our
solar-system. Finding out, what the nature of all these planets are,
is one of the most interesting questions of exoplanet research.

Studying hot Jupiters is all but boring, as we do not really know how
they form. Do they form via disk-induced migration, high-eccentricity
migration, or in in-situ? The discovery of WASP-107\,b, a planet mass
in the Neptune regime and a radius of Jupiter, can best be explained
with a formation at large distance with subsequent migration
\cite{Piaulet2021}. However, WASP-47, a system with a hot Jupiter
sandwiched between an inner and an outer low-mass planet favors
in-situ formation \cite{Batygin2016}. It thus looks like that there is
more than one path-way of planet formation \cite{Hallatt2020}.  To
unravel this mystery, we have to determine the masses, and radii of
many exoplanets accompanied by studied of their atmospheres.

The evolution of planetary atmospheres is a complicated process which
involves many factors.  One of the most important factors is the
amount of XUV-radiation (X-ray + EUV) that the planets receive.  The
XUV-radiation is the driving force for the photochemistry of planetary
atmospheres and it is likely to play a key role for the erosion of
planetary atmospheres.  As we will discuss in the section \ref{sec:2},
UV-lines can be used as tracers for the XUV-radiation from the host
stars which can otherwise only be studied by satellites.

The big leap forward in exoplanet research will be the exoplanet
characterization mission PLATO (PLAnetary Transits and Oscillations of
stars)\cite{Rauer2014}, and the atmospheric characterization mission
ARIEL (Atmospheric Remote-sensing Infrared Exoplanet
Large-survey)\cite{Puig2018}.  PLATO is foreseen to be launched in
2026, and ARIEL in 2029.

PLATO will use the transit-method to find planets down to the size of
the Earth orbiting, solar-like stars at distances up to the habitable
zones, and determine their radii with extraordinary high precision.
The payload consists of 24 'normal' and 2 'fast' cameras.  The fast
cameras will be used to monitor stars of $\rm 4 \leq V \leq 8$
mag. The prime targets of the normal cameras will be stars brighter
than $\rm V\leq 11$\,mag. It is estimated that PLATO will discover
4000 super-Earths and Earths , of which 40-70 will be in the habitable
zones.  The aim is to determine the masses of these planets using the
RV-method. As we will explain in the Section \ref{sec:3}, this is all
but easy because of the stellar activity.  As already mentioned,
stellar activity also plays an important role for the mass-loss and
photochemistry of planetary atmospheres. It is also of key importance
for the existence of life \cite{Abrevaya2020}. Stellar activity thus
has to be studied for all planet host stars, particularly for those
that harbor potentially habitable planets.

ARIEL is the first space mission dedicated to measuring the chemical
composition and thermal structures of hundreds of transiting
exoplanets. ARIEL has three photometric channels in the VIS (0.5-0.6
$\rm \mu m$, 0.6-0.81 $\rm \mu m$, 0.81-1.0 $\rm \mu m$) and three
spectroscopic channels in the NIR (1.1-1.95 $\rm \mu m$, R=20;
1.95-3.9 $\rm \mu m$, R=100, 3.9-7.8 $\rm \mu m$, R=30). ARIEL will
observe 1000 preselected transiting planets, of which 50-100 will be
studied intensively.  Since the signal from the atmosphere is
proportional to radius of the planet times the scale height, the prime
targets are planets with large scale heights. The scale height is
given by $\rm H=k_B\,T/(\mu g)$, with $\rm k_B$ the Boltzmann's
constant, T the temperature of the atmosphere, $\rm \mu$ the mean
molecular mass, and g the surface gravity of the planet
\cite{Griffith2014}. Most ARIEL targets will have temperatures in the
range between 1000 to 2000 K. More than half of them will be larger
than $\rm 4\,R_{Earth}$. The high temperature of the targets not only
increases the signal-to-noise ratio of the spectra, it also ensures
that the atmospheres are well mixed.  In summary, most of the ARIEL
targets will be close-in, relatively large planets orbiting relatively
bright stars.

One of the experiments that will be carried out by ARIEL is the
determination of the C/O-ratio in hot Jupiters.  $\rm C/O\sim1$ would
indicated that the planet has formed beyond the snow line, and $\rm
C/O\sim0.5$ would indicate a formation inside the ice-line
\cite{Oeberg2011}. However close-in planets receive a lot of
XUV/UV-radiation which drives the photochemistry in planetary
atmospheres.  We thus need to study the XUV- and UV-radiation as well,
at least indirectly.

We also need to measure the masses of the planets, because that
determines the scale height. What we also need are the element
abundances of the host star to be compared with the abundances of the
planets \cite{Sheppard2021}. Because the RV-variations of hot/warm
Jupiter- and a Neptune-mass planets are 200-400, and 10-22 $\rm
m\,s^{-1}$, respectively, we do not need an extreme RV-precision.  In
summary, what we need for PLATO and ARIEL is the possibility to study
stellar activity, the possibility to determine stellar parameters and
an RV-precision of 10 $\rm m\,s^{-1}$, or better.

\section{The Ca\,{\tiny II}\,H\&K and other UV-lines 
tracers for the XUV-radiation}
\label{sec:2}

For understanding the evolution of planets, it is essential to
understand the erosion processes of planetary atmospheres. While
different processes have been suggested, atmospheric losses due to the
XUV-radiation from the host ($\rm \lambda < 91.2\,nm$) certainly plays
an important role \cite{Lammer2014}, \cite{Linsky2015}. While the main
erosion phase for planets of solar-like stars is during the first 100
Myrs, extreme mass-losses have also been observed on planets orbiting
stars older than that, for example WASP\,12b \cite{Hebb2009} and
WASP-121\,b \cite{Delrez2016}. Studying the activity of stars hosting
planets with extreme mass-loss is the best way to learn more about
these processes.

Unfortunately, the wavelength region $\rm \lambda < 91.2\,nm$ can only
be observed from space. However, Sreejith et al.  \cite{Sreejith2020}
showed that the amount of X-ray and EUV-radiation correlates well with
the Ca\,{\tiny II}\,H\&K flux.  The flux of the star at XUV-wavelength
can thus be estimated from the flux of the Ca\,{\tiny II}\,H\&K lines.

Flares contribute to 20-50\% of XVU-flux in young stars
\cite{Guedel2004}. It is thus important to study the flare activity
for understanding atmospheric erosion and photochemistry of planetary
atmospheres.  The contribution from flares is critical, because the
temperatures of flares are higher than that of the corona. That means,
the XUV-spectrum of the flares is harder, and the photons penetrate
deeper into the atmospheres of the planets. The basic properties of
the flares can be determined from diagnostic emission lines, most of
which are in the UV-regime that can still be observed by ground-based
telescope \cite{Fuhrmeister2018}.  Because active stars have a large
number of small flares, the lines from flares are present most of the
time. Monitoring of such
flares is one of the cases being developed for the new CUBES instrument
for the VLT (see Zanutta et al, this volume).  However, for observations
of the CaII\,H\&K lines (393 \& 397nm) in bright stars a smaller telescope than
the VLT can suffice. In Section~\ref{sec:4}. we present plans for the new PLATOspec
instrument for the 1.5m telescope at La Silla, that will neatly complement
future CUBES observations of fainter, more demanding targets.

\section{Stellar activity and the RV-jitter}
\label{sec:3}


As already outlined in Section~\ref{sec:1}, we have to determine not
only the radii but also the masses of exoplanets. Ideally the masses
should be determined by using the RV-method, rather than
TTVs. Numerical simulations by \cite{Cumming2009} showed that 100
RV-measurements allow to detect planets with K-amplitudes that are 1.5
(1) times larger than the noise-level of the measurement.  After
carrying out the RV-fitting challenge Dumusque et al.
\cite{Dumusque2017} conclude that $\rm K/N>7.5$ gives a 80-90\%
recovery rate, where $\rm K/N=K_{pl}/RV_{rms}\times \sqrt{N_{obs}}$.
This means, for N=100, $\rm K_{pl}\geq 0.75\,RV_{rms}$.  If we aim for
the detection of a planet with 10(4) $\rm M_{Earth}$ at one AU, we
require 100 RV-measurements with an accuracy better than 1.2 (0.5)
$\rm m\,s^{-1}$.

Instruments like CARMENES, HARPS HAPS-N, routinely achieve an
RV-accuracy of about one $\rm m\,s^{-1}$, and the design-goal of
ESPRESSO is an RV-accuracy of 0.1 $\rm m\,s^{-1}$. However, the
RV-accuracy does not only depend on the precision of the instrument,
it also depends on the activity-level of the star.  Su{\'a}rez
Mascare{\~n}o et al.  \cite{Suarez Mascareno2017} have studied the
relation between Ca\,{\tiny II}\,H\&K-index and the K-amplitude of the
activity.  For a stellar-noise jitter of 0.4-0.5 $\rm m\,s^{-1}$, a
GK-star has to have $\rm log_{10}(R'_{HK})\sim -5.0$. This is about
the current solar-activity level. According to Hall et
al. \cite{Hall2007} 32\% of the solar-like stars have this activity
level, or are even less active.  It should thus be possible to detect
planets with $\rm M_{p} \sim 4\,M_{Earth}$ in the habitable zone of
solar-like stars if the star also has a solar-like activity level.
Collier Cameron et al. \cite{CollierCameron2019},
\cite{CollierCameron2021} made a study of the sun as a star using a
time-series of 853 daily observations with a $\rm 250 < S/N < 400$
obtained with HARPS-N. They injected synthetic low-mass planet signals
corresponding to planets of 1.2, 1.9, 2.9, and 3.7 $\rm M_{Earth}$
with periods of 7.142, 27.123, 101.543, and 213.593 d. Using a
sophisticated method that allows to remove the stellar activity, they
demonstrated that such planets could be detected.  The results from
Collier Cameron et al.  \cite{CollierCameron2021} thus agree well with
previous estimates \cite{Cumming2009},\cite{Dumusque2017}.

K-stars are even better targets than G-stars, because the
RV-amplitudes caused by planets in the habitable zones are larger. In
this case, the limit is about 2 $\rm M_{Earth}$. In summary, planets
with about 2-4 $\rm M_{Earth}$ can be detected orbiting solar like
stars if their activity level is low.  It is thus important to monitor
the PLATO targets extensively in Ca\,{\tiny II}\,H\&K before
attempting to determine the masses of the planets.

\section{PLATOspec}
\label{sec:4}
\subsection{Science goal and requirements for the instrument}
\label{sec:4.1}

\begin{figure*}
  \includegraphics[width=0.75\textwidth]{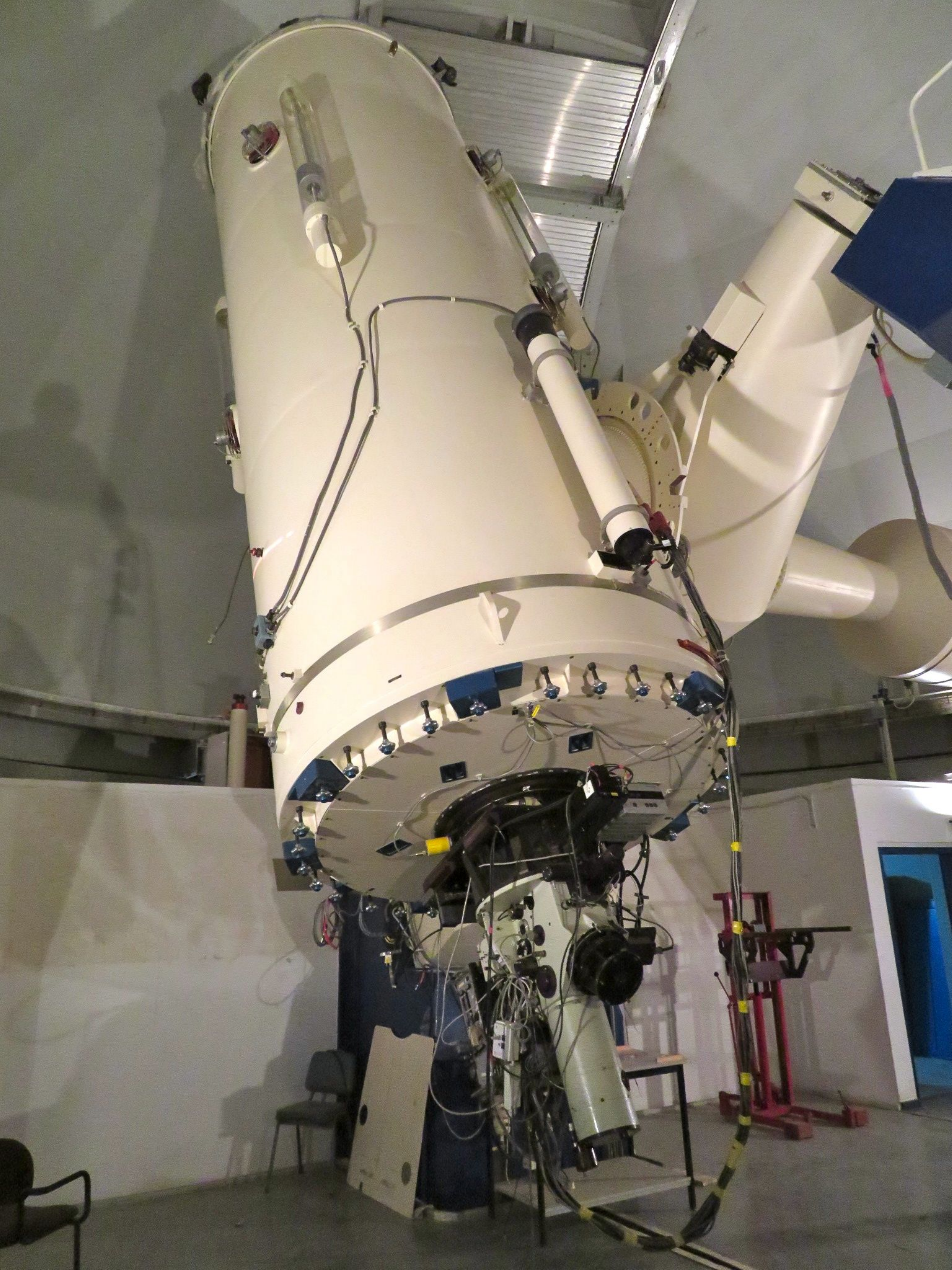}
 \caption{The ESO 1.5m telescope}
 \label{fig:1}    
 \end{figure*}

\begin{figure*}
  \includegraphics[width=0.75\textwidth]{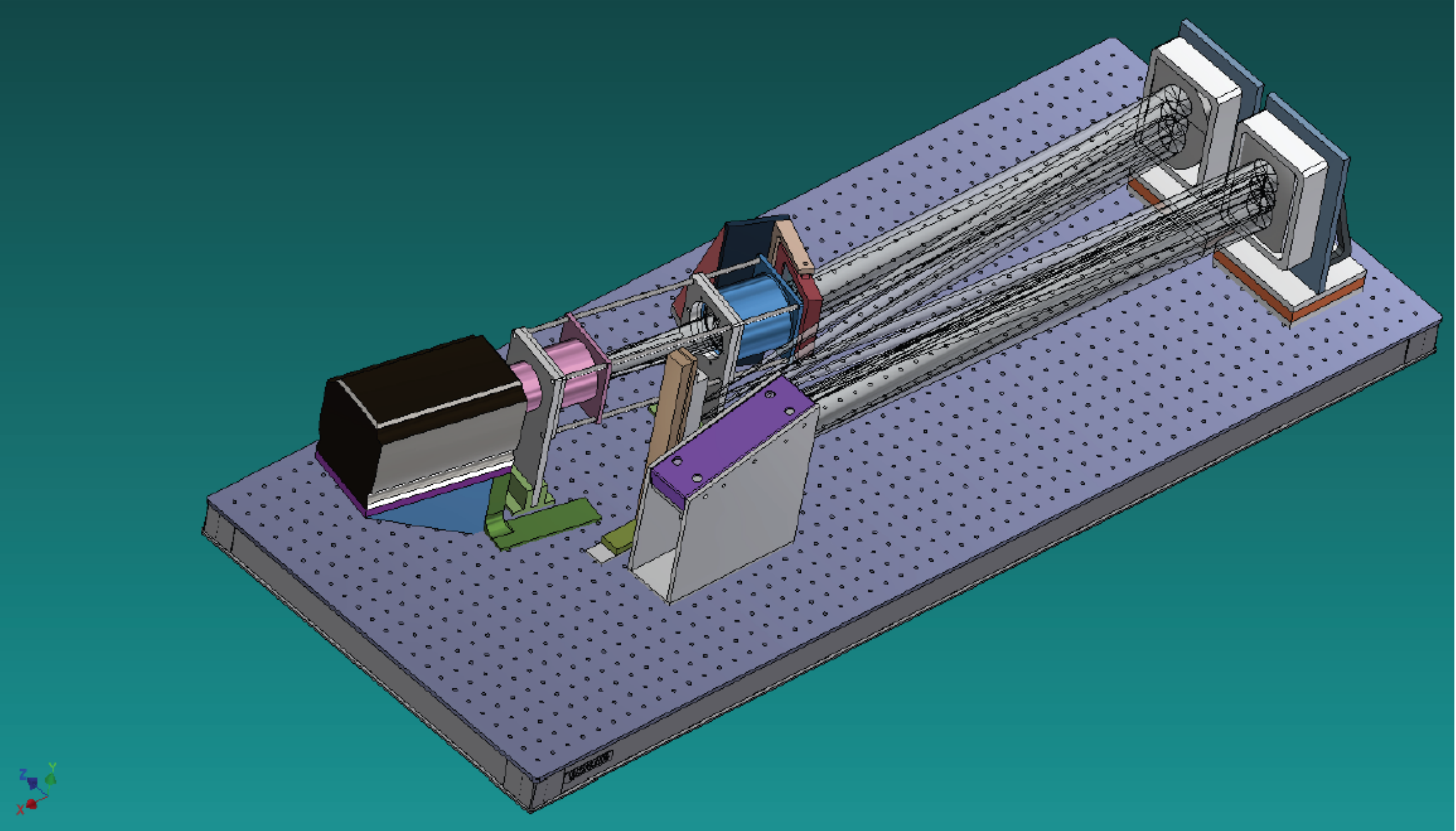}
 \caption{Preliminary mechanical lay out of the PLATOSpec spectrograph.}
 \label{fig:2}    
 \end{figure*}

 \begin{figure*}
  \includegraphics[width=0.50\textwidth]{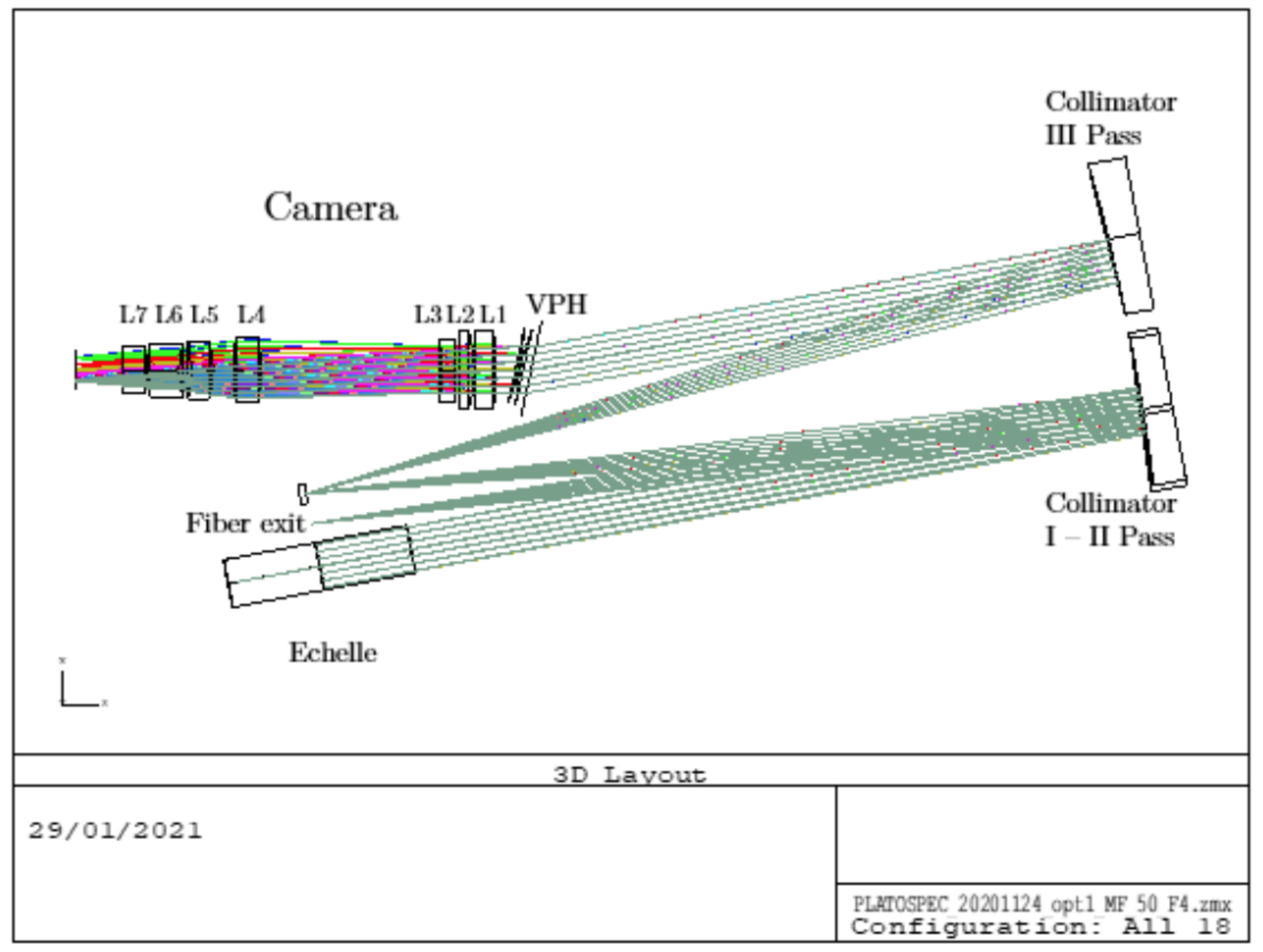}
 \caption{Spectrograph optical layout in WP configuration. .}
 \label{fig:3}    
 \end{figure*}

As we have outlined in sections ~\ref{sec:1},\ref{sec:2},\ref{sec:3},
PLATO and ARIEL will be the next big steps in exoplanet research.
However, for these missions we need extensive ground-based support
observations. Amongst these are extensive studies of the Ca\,{\tiny
  II}\,H\&K lines and RV-measurements with an accuracy of 10 $\rm
m\,s^{-1}$, or better. Given that PLATO will discover 4000 planets
that calls for a dedicate instrument just for these observational
needs. We thus initiated the PLATOspec project, which has three
scientific aims:

\begin{itemize}
\item[i.)] Measure the flux in the Ca II H\&K lines of
  planet host stars.
\item[ii.)]  Determine the stellar parameters Teff, log(g) abundance,
  and space velocity of planet host stars.
\item[iii.)] Determine the masses of hot and warm Jupiters, and hot
  Neptunes.
\end{itemize}

Requirements of the instrument: 

\begin{itemize}
\item[1.)] High efficiency at a wavelength of 390 nm.
\item[2.)] Location at suitable site: The instrument needs to be
  located at a site with more than 2000 hours of suitable observing
  conditions which also has a high sky-transmission in the UV.
\item[3.)] Resolution and wavelength coverage: $R\sim70000$ (4.2 $\rm
  km\,s^{-1}$) and a wavelength coverage of 360-680 nm.
\item[4.)] RV-accuracy: A quick scan mode with an RV-accuracy better
  than 10 $\rm m\,s^{-1}$ for stars down to V=11 mag, and a high-precision 
  mode with an RV-accuracy of 3 $\rm m\,s^{-1}$ for special targets. 
\end{itemize}

\subsection{The instrument}
\label{sec:4.2}

PLATOspec is the new high-resolution high-stable spectrograph for the
ESO 1.5m telescope of the Observatory La Silla (Fig.~\ref{fig:1}). The
Project is lead by the Czech Academy of Science (PI Petr Kabath) in
equal share with two main partner institutes, the Observatory of
Tautenburg (CoPI Artie Hatzes) and the Center of Astro Engineering UC
(CoPI Leo Vanzi). In addition the Masaryk University from the Czech
Republic and the Universidad Adolfo Ibanez from Chile join the
project as minor partners in 2021.

PLATOspec will be a fibre-fed Echelle spectrograph that will be placed in a
climatized room.  The telescope is being refurbished and upgraded in
2021.  The focal ratio of the telescope is F/14.9 which provides a
scale of about 9.2 arcsec/mm at its focal plane. This allows
projecting an aperture of about 2 arcsec diameter of the sky onto a 50
$\rm \mu m$ core fiber. In particular 1.8 arcsec at F/3.8 or 2.0
arcsec at F/3.4 with suitable feeding optics. An octagonal fiber can
be used for optimal scrambling of the near field illumination.

The instrument concept follows similar instruments of the same class
such as FEROS \cite{Kaufer1998}, CORALIE \cite{Queloz2000} or CHIRON
\cite{Tokovinin2013} (Fig.~\ref{fig:2},\ref{fig:3}).  In particular
our team will use as reference the spectrograph FIDEOS
\cite{Vanzi2018} and TCES the Echelle spectrograph of the 2-m-Alfred
Jensch telescope at Th\"uringer Landessternwarte Tautenburg because of
our direct experience with them.  TCES has $R\sim 67000$, and an
iodine cell (IC).  For very bright stars, we have achieved an accuracy
of 1.2 to 1.7 $\rm m\,s^{-1}$ with a similar instrument
\cite{Hatzes2007}.

There will be two mode for precision RV-measurements:

1.) The medium precision RV-mode: In this mode the simultaneous
calibration technique with an ThAr hollow-cathode lamp (HCL) is used.
PLATOspec will not be evacuated, only temperature stabilized to a
level of about 0.1 C. Experience with similar instruments show that an
accuracy of 8 $\rm m\,s^{-1}$ can be achieved.  In particular the
results of FIDEOS indicates that an RV precision of 4-5 $\rm
m\,s^{-1}$ is possible. A further improvement of this mode are
possible.  Experiments with TCES shows that the main shifts are caused
by changes of the pressure. By including a device to measure the
pressure changes during the night and using this information in the
data-pipeline, the drifts can be minimized.

2.) A high precisions RV-mode: In this mode, an iodine cell (IC )is
inserted in to the beam so that the star-light goes through it. The IC
provides a very dense grid of lines between 510 nm to 620 nm. The IC
will be used for modeling precisely the drift of the spectrograph and
the changes of the point spread function. Since only small part of the
spectrum is used for the RV-measurements and 30\% of the light is
absorbed in this case, this mode is recommended only for relatively
bright stars. We expect that RV-precision of about $\rm 3\,m\,s^{-1}$,
similar to TCES, will be achieved.

We expect that the instrument will become operational in 2023.

\begin{acknowledgements}
This work was generously supported by the Th\"uringer Ministerium
f\"ur Wirtschaft, Wissenschaft und Digitale Gesellschaft and by the
Th\"uringer Aufbaubank. P.K. acknowledges the support of MSMT 
grant no. LTT-20015.

\end{acknowledgements}

%
%



\end{document}